\documentclass[journal,final, 10pt]{IEEEtran}
\IEEEoverridecommandlockouts
\usepackage{cite}
\usepackage{amsmath,amssymb,amsfonts}
\usepackage{algorithmic}
\usepackage{graphicx}
\usepackage{textcomp}
\usepackage{xcolor}
\usepackage{multirow}%专门处理表格跨列
\usepackage{makecell}%专门处理表格跨列
\def\BibTeX{{\rm B\kern-.05em{\sc i\kern-.025em b}\kern-.08em
    T\kern-.1667em\lower.7ex\hbox{E}\kern-.125emX}}
\usepackage{amsmath}
\usepackage{array}
\usepackage{longtable}
\usepackage{threeparttable}
\usepackage{booktabs}
\usepackage{pifont}
\begin{document}

\title{Foresight-Seeing Quantum Reinforcement Learning for Two-Stage Unit Commitment with Virtual Power Plants and Renewable Power Integration
}

\author{Xiang Wei,~\IEEEmembership{Student Member,~IEEE,} Ziqing Zhu,~\IEEEmembership{Member,~IEEE,} Linghua Zhu, Ze Hu, Xian Zhang,~\IEEEmembership{Member,~IEEE,} Guibin Wang,~\IEEEmembership{Member,~IEEE,}  Siqi Bu,~\IEEEmembership{Senior Member,~IEEE}, Ka Wing Chan,~\IEEEmembership{Member,~IEEE} 
        % <-this % stops a space
}

\maketitle

\begin{abstract}
Unit commitment (UC) optimizes the start-up and shutdown schedules of generating units to meet load demand while minimizing costs. However, the increasing integration of renewable energy introduces uncertainties for real-time scheduling. Existing solutions face limitations both in modeling and algorithmic design. At the modeling level, they fail to incorporate widely adopted virtual power plants (VPPs) as flexibility resources, missing the opportunity to proactively mitigate potential real-time imbalances or ramping constraints through foresight-seeing decision-making. At the algorithmic level, existing probabilistic optimization, multi-stage approaches, and machine learning, face challenges in computational complexity and adaptability. To address these challenges, this study proposes a novel two-stage UC framework that incorporates foresight-seeing sequential decision-making in both day-ahead and real-time scheduling, leveraging VPPs as flexibility resources to proactively reserve capacity and ramping flexibility for upcoming renewable energy uncertainties over several hours. In particular, we develop quantum reinforcement learning (QRL) algorithms that integrate the foresight-seeing sequential decision-making and scalable computation advantages of deep reinforcement learning (DRL) with the parallel and high-efficiency search capabilities of quantum computing. Experimental results demonstrate that the proposed QRL-based approach outperforms in computational efficiency, real-time responsiveness, and solution quality. 
\end{abstract}

\begin{IEEEkeywords}
Unit Commitment, Quantum Reinforcement Learning, virtual power plants, Quantum Markov Decision Process
\end{IEEEkeywords}

\vspace{-0.6cm}
\section{Introduction}

%机组组合（Unit Commitment, UC）是电力系统优化调度的核心问题之一，主要解决在给定的负荷需求下，如何在一定的时间范围内选择合适的发电机组启动和停运，并在后续决定它们的出力计划以满足电力需求。机组组合的核心目标是在确保电力系统安全稳定运行的前提下，最小化系统的运行成本，包括燃料成本、启停机成本和排放成本等。 但是，随着电力系统中新能源发电所占的比重不断提升，其出力不确定性和快速波动特性给电力系统的实时运行带来了巨大的挑战。举例来说，由于太阳能发电量的剧烈波动，电力需求曲线在一天内呈现出一种类似鸭子形状的变化。白天，太阳能发电减少了对常规发电机组的依赖，电网负荷曲线被压低；然而，随着太阳能发电量的下降，特别是傍晚时分，电力需求急剧上升，这时系统需要具有快速增加发电量的能力（即向上调节能力），以弥补突然减少的太阳能输出。再或者，随着天气的突然变化，风速的突然加大会造成风电场的出力突然急剧增加，此时系统需要具有快速减少总发电出力的能力（即向下调节能力）。这种情况下，传统的基于静态时间尺度（如小时或更长时间）的机组组合方法显得不足以应对复杂的系统需求，因此其范式需要做重大的调整来适应这种变化。

% 第一种调整的思路是前瞻性的动态规划（Foresight Dynamic Programming），即对于每个时段的决策，都提前考虑未来几个时刻的发电需求和新能源发电量的变化趋势，从而提前在决策模型中预留出足够的“爬坡”能力。这种前瞻性可以体现在日内的实时经济调度上，也可以体现在日前机组组合上，
%这也意味着日前-日内的两阶段决策过程的协调变得非常重要。（介绍方法思路+review文献） (两阶段日内日前 UC problem,)

% 第二种调整的思路是利用概率性预测，进行基于新能源不确定性概率分布不确定集的鲁棒/分布鲁棒优化。（介绍方法思路+review文献）

% 然而，无论是前瞻性动态规划还是鲁棒/分布鲁棒优化，其均给电力系统机组组合问题的求解带来了巨大的挑战。（分析为什么有挑战，引出机器学习方法（包括端对端的深度学习、机器学习增强的传统优化方法，即用机器学习方法求解传统优化当中最难的一些步骤），分别讨论其优缺点）

% 近年来，强化学习的发展为求解两阶段机组组合问题提供了新的思路。（介绍强化学习+为什么能够解决上边的方法解决不了的问题，之后review相关文献，最后讨论强化学习的缺点，同样参考下边这个链接，要和QRL优点一一对应）

% 量子计算的优势：可参考\href{https://eitca.org/artificial-intelligence/eitc-ai-tfqml-tensorflow-quantum-machine-learning/quantum-reinforcement-learning/replicating-reinforcement-learning-with-quantum-variational-circuits-with-tfq/examination-review-replicating-reinforcement-learning-with-quantum-variational-circuits-with-tfq/what-are-the-potential-advantages-of-using-quantum-reinforcement-learning-with-tensorflow-quantum-compared-to-traditional-reinforcement-learning-methods/}{What are the potential advantages of using quantum reinforcement learning with TensorFlow Quantum compared to traditional reinforcement learning methods? - EITCA Academy} 

% 本文最重要的创新点：QRL = 前瞻性的序贯决策 + 数据表征能力

\subsection{General Background}
Unit commitment (UC) is a key challenge in power system operation \cite{2.1}, focusing on selecting generating units to start-up or shut down within a specified time to meet load demand. The goal is to minimize operational costs, including fuel, start-up/shutdown, and emissions costs, while ensuring system stability. However, the growing share of renewable energy introduces uncertainties and fluctuations that challenge real-time operations \cite{1}. For example, solar power creates a "duck curve", with reduced daytime demand but sharp evening increases requiring rapid generation ramp-up \cite{2.2, 2.3}. Similarly, sudden weather changes, like rapid wind speed increases, can cause wind power surges, demanding swift downward regulation to maintain balance \cite{2.4}. Hence, traditional UC methods with static time scales, such as hourly intervals, are inadequate for these dynamic needs \cite{2.5}, necessitating a paradigm shift in UC methodologies to address these challenges.
\vspace{-0.4cm}

\subsection{Literature Review}
The first adjustment approach uses the probabilistic distribution of renewable energy output with stochastic \cite{1.5}, robust \cite{1.6}, or distributionally robust optimization techniques \cite{1.8}. These methods require high accuracy in probabilistic distributions and involve significant computational complexity. The second approach integrates two- or three-stage optimization, combining day-ahead UC, intra-day UC, and real-time ED \cite{1.1,1.2,1.3,1.4}. It employs foresight dynamic programming to predict power demand and renewable energy fluctuations, ensuring reserved "ramping" capability in advance \cite{1.1,1.2}. Although methods like lagrangian relaxation \cite{1.3} or distributed optimization \cite{1.4} reduce computational complexity by decomposing problems, foresight dynamic programming significantly increases computational burden by expanding the search space to account for current and future states. Nowadays, advances in AI and machine learning have significantly accelerated UC problem-solving through end-to-end approaches \cite{1.9,1.10,1.11,1.12} and integration with traditional optimization algorithms \cite{9388933,10026495}.  However, machine learning-based UC decision-making faces key challenges: Firstly, deep learning, as a data-driven fitting method, lacks real-time adaptability to changes in power system states, limiting its effectiveness in systems with abundant renewable energy \cite{1.14}. Secondly, deep learning relies on high-quality training data; inadequate data can lead to underfitting or overfitting, reducing accuracy and generalizability \cite{1.15}.

Compared to traditional optimization and machine learning methods, deep reinforcement learning (DRL) offers unique advantages for solving the UC problem. Firstly, DRL utilizes Markov decision processes (MDP), incorporating dynamic programming principles to optimize sequential decision-making under uncertainty \cite{2.6}. This allows DRL-based UC scheduling to predict generator statuses and reserve ramping capacity to handle uncertainties \cite{19}. Secondly, DRL minimizes reliance on pre-existing data by employing interactive learning with the environment. Through iterative training, it continuously refines strategies to maximize cumulative rewards \cite{1.18}. Finally, DRL leverages deep learning to approximate complex value functions and policies. Unlike traditional methods, it emphasizes sequential decision-making strategies across multiple time steps, enabling adaptable solutions \cite{2.8}. Furthermore, robust or distributionally robust programming can be integrated into DRL frameworks, such as robust RL, to address operational uncertainties.

%\begin{table*}[!ht]
%\centering
%\caption{Computational Performance Comparison among Different Methods}
%\vspace{-0.2cm}
%\label{tab:optimization_requirements}
% %\parbox[t]{2cm}{Infeasible in the face of uncertainties}
%\begin{tabular}{l c c c c c}
%\toprule
%\multirow{3}{*} {Methods} & \multicolumn{5}{c}{\textbf{Optimization requirements}} \\ \cmidrule{2-6}
%                 & \parbox[t]{2cm}{Computation Speed} &\multirow{2}{*} 
% {Scalability} & \parbox[t]{2cm}{Online Deployment}  & \multirow{2}{*} {Modeling Requirements} & %\multirow{2}{*} {Optimality Guarantee}\\ \midrule
%\multirow{3}{*} {\parbox[t]{2cm}{Mathematical programming}}
%    & \multirow{3}{*} \times& \multirow{3}{*} \times& \parbox[t]{2.5cm}{Cannot adjust promptly to %real-time changes}  & \parbox[t]{2.5cm}{Rely on precise mathematical models}&\checkmark\\
%    
%\multirow{2}{*} {Machine learning} & \multirow{2}{*} \checkmark & \multirow{2}{*} \checkmark& %\parbox[t]{2.5cm}{Rely on pre-learned static patterns}  & \parbox[t]{2.5cm}{Need high-quality %training data}& \times\\
%
%\multirow{2}{*} {Classic RL}            & \multirow{2}{*} \checkmark & \multirow{2}{*} \checkmark & %\multirow{2}{*} \checkmark & \parbox[t]{2.5cm}{Depend on sufficient exploration}& \parbox[t]{2.8cm}%{Face local minima and saddle points problem}\\
%
%QRL               & \checkmark & \checkmark & \checkmark & \checkmark & \checkmark\\ 
%\bottomrule
%end{tabular}
%\end{table*}
%\vspace{-0.4cm}
%\vspace{-0.4cm}

\begin{table*}[!ht]
\centering
\renewcommand{\arraystretch}{1.3}
\caption{Computational Performance Comparison among Different Methods}
\label{tab:optimization_requirements}
\begin{tabular}{c c c >{\centering\arraybackslash}p{3cm} >{\centering\arraybackslash}p{3cm} >{\centering\arraybackslash}p{3cm}}
\toprule
\multirow{2}{*}{\textbf{Methods}} & \multicolumn{5}{c}{\textbf{Optimization Requirements}} \\ \cmidrule{2-6}
 & Computation Speed & Scalability & Online Deployment & Modeling Requirements & Optimality Guarantee \\ 
\midrule
Mathematical Programming & × & × & Cannot adjust promptly to real-time changes & Rely on precise mathematical models & \checkmark \\
Machine Learning & \checkmark & \checkmark & \checkmark (but need high-quality training data) & No specific requirements & × \\
Classic RL & \checkmark & \checkmark & \checkmark & No specific requirements & Face local minima and saddle points problem \\
QRL & \checkmark & \checkmark & \checkmark & \checkmark & \checkmark \\ 
\bottomrule
\end{tabular}
\end{table*}
\vspace{-0.4cm}

\subsection{Motivations and Main Contributions}
Based on the literature review, we identify the following shortcomings in existing solutions to the UC problem: Firstly, in terms of modeling, the integration of distributed generation has made virtual power plants (VPPs) a vital component of modern power systems \cite{3.1}. VPPs provide flexible regulation and effectively complement flexibility resources in two-stage UC frameworks, enhancing the main grid's ramping capacity \cite{3.2}. However, existing literature lacks successful implementations of VPPs as flexibility resources in two-stage UC models. Secondly, in terms of algorithms, while DRL overcomes many limitations of traditional methods, it has significant drawbacks as a search algorithm. The agent must evaluate numerous potential actions under varying power system states \cite{2.6} (e.g., different load levels) and balance exploration \cite{3.4} (discovering better actions) with exploitation (maximizing known rewards). Finally, DRL algorithms rely on gradient-based optimization, which is prone to issues like local minima and saddle points \cite{3.5}. These challenges, particularly in large-scale UC problems and two-stage optimization models, can degrade solution quality.

Recent advancements in quantum computing offer innovative solutions to power system optimization problems \cite{3.6}. Quantum computing uses qubits' unique properties of superposition and entanglement, enabling them to exist in multiple states simultaneously \cite{3.7}. This allows quantum algorithms to process large amounts of information in parallel, potentially achieving exponential speedups compared to classical methods \cite{3.8}. Specifically, superposition enables exploration of multiple states simultaneously, enhancing efficiency in discovering optimal strategies. Additionally, quantum variational circuits use parameterized quantum gates to represent policies or value functions in reinforcement learning \cite{3.9}, with parameters optimized through classical techniques. The quantum properties of these circuits allow broader exploration of the solution space, reducing the risk of local minima and increasing the likelihood of identifying superior global optima.

Inspired by the above research gaps and recent advancements, this work proposes a two-stage UC framework leveraging distributed VPPs as flexibility resources, while we innovatively synergize quantum computing and DRL to improve the computational performance. Key contributions of this work are summarized as follows:
 
1) We develop a novel two-stage UC optimization model characterized with two significant features: multi-timescale foresight-seeing optimization and the integration of VPPs as a novel flexibility resource. In the day-ahead stage, the model employs foresight dynamic programming (DP) to optimize the startup and shutdown schedules and power allocation of traditional generators, while reserving adequate ramping capacity to prepare for potential fluctuations. In the real-time stage, the model leverages the dynamic adjustment capabilities of VPPs to precisely compensate for power imbalances arising from uncertainties.

2) We re-formulate the two-stage UC model as MDP, which is quantumized by developing an innovative quantum Markov decision process (q-MDP) framework. In this framework, classical state-action probability distributions are replaced with quantum density operators, and system state transitions are described using quantum channels. Additionally, we reformulate the UC optimization objective into a quantum reward function using the Hilbert-Schmidt inner product and quantum observables. This innovative approach lays theoretical foundation to leverage quantum advantages for computational performance enhancement.

3) We develop two quantum reinforcement learning (QRL) algorithms to solve the q-MDP. First, we introduced parameterized quantum circuits (PQC) as Q-function approximators to replace traditional deep neural networks (DNNs), utilizing quantum superposition and entanglement to enhance state representation capabilities. Leveraging the PQC, we quantumized classical deep Q-network (DQN) and soft actor-critic (SAC) algorithms with superior performance in convergence and optimality. Table I provides a detailed comparison of the proposed QRL with existing methods in solving UC problems.

The rest of this work is organized as follows. Section II introduces a foresight-seeing UC model. The framework of the q-MDP model is described in section III, and section IV presents the basic principles of QRL. Simulation results are demonstrated in section V, and section VI concludes this work.

%\begin{table}[!ht]	
%	\centering
%	\caption{Training results with different algorithms}\label{tab:1}
%	\vspace{-0.25cm}
%	\begin{tabular}{c|cccccc}
%		\toprule
%		Methods& [3] & [9] & [16] & [19] &[20] &[21] \\ \midrule
%		Mathematical programming& * &  & *&  &* &\\
%		Machine learning&  & * & &*  & & \\
 %       Classic RL&  & & * & *& *&* \\
  %      QRL&  & &  & & & *\\
	%	\bottomrule
	%\end{tabular}
%\end{table}

\begin{figure*}
	\centering
	\includegraphics[width=0.8\textwidth]{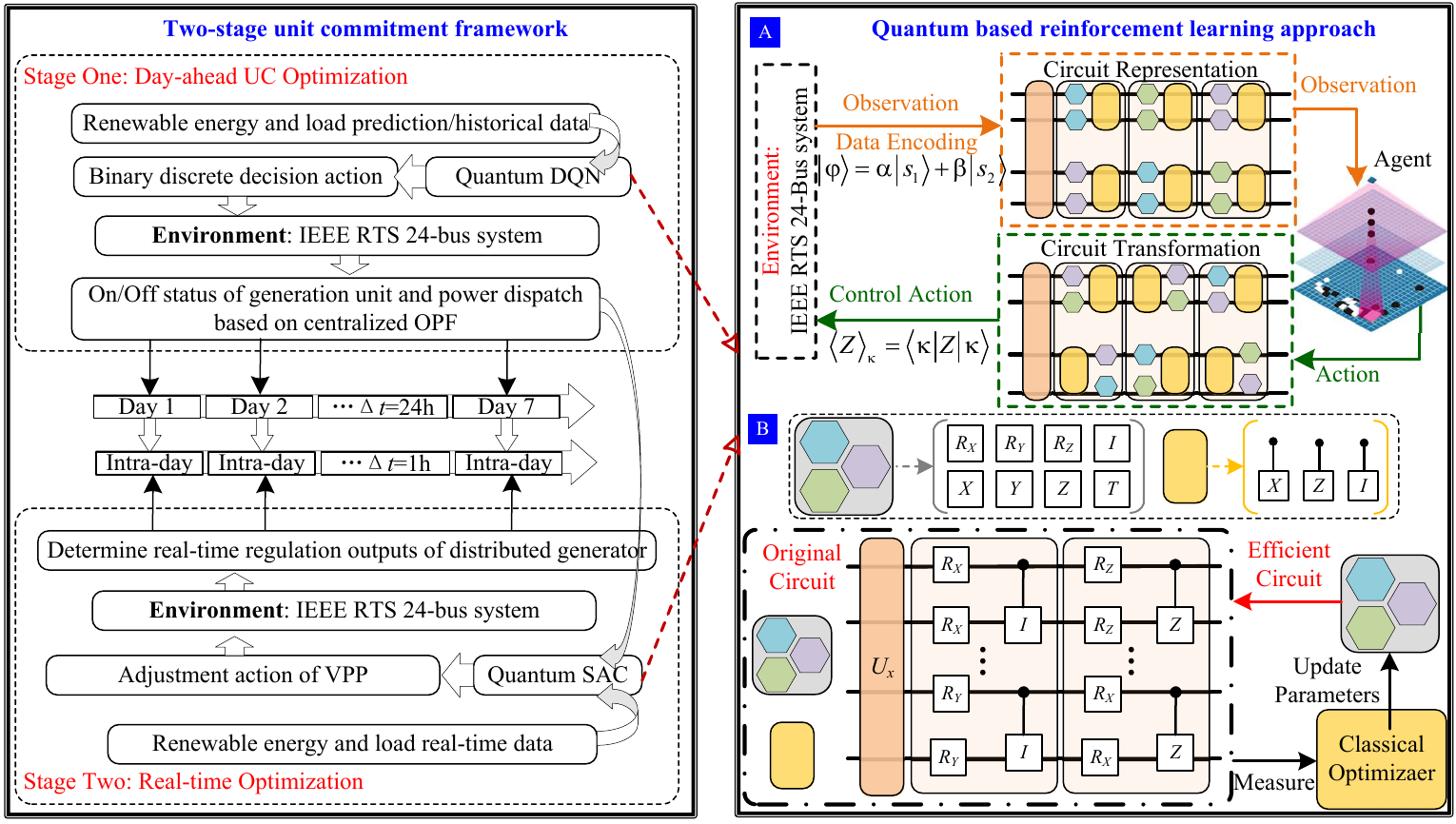} % Use textwidth for spanning both columns
	\vspace{-4mm}
	\caption{Proposed two-stage robust UC Framework. A) The quantum-based RL approach encodes observational data into a quantum circuit for optimization. Using a diagrammatic representation of the circuit, the agent selects circuit transformations to generate control actions, which are then decoded and applied to the environment. This process is repeated iteratively. B) All possible single-qubit gates are represented by hexagons, while two-qubit gates are denoted by yellow rectangles at the top. The unitary \( U_x \), shown below, corresponds to the data encoding layer. The quantum circuit iteratively optimizes trainable parameters for all candidate qubit gates based on a gradient-based classical optimizer, transforming the original circuit into a more efficient configuration. }
	\label{fig:1}
	\vspace{-0.4cm}
\end{figure*}

\section{Problem Formulation: Two-Stage Foresight-Seeing UC with Flexible VPPs}
In this section, we introduce an innovative two-stage UC model that coordinates decisions across multiple timescales and integrates VPPs as flexible resources, as shown in Fig. 1 (left-hand side). In the day-ahead stage, the model applies foresight DP to optimize generator startup/shutdown schedules and power allocations, ensuring sufficient ramping capacity to accommodate potential fluctuations. In the real-time stage, the model continues to leverage foresight DP by dynamically adjusting VPP dispatch to proactively mitigate imbalances caused by renewable energy variability and demand uncertainties. 
\vspace{-0.2cm}

\subsection{Day-ahead Foresight-Seeing Dynamic Programming}
The goal of day-ahead UC dispatch is to determine the most cost-effective UC schedule and power generation allocation that satisfies the load demand over a specific time period while adhering to a set of operational constraints. The detailed formulation is given as follows:
\vspace{-0.6cm}

\begin{align}
&\min \sum_{t=1}^T \sum_{g} C_g^{su} + \sum_{t=1}^T \sum_{g} C_g P_{g,t} + \sum_{t=1}^T \sum_{n} C_{ls} P_{n,t}^{LS} \tag{1}\\
\text{s.t. } &\quad P_{n,g,t} + P_{n,pv,t}^{DA} + P_{n,w,t}^{DA} = \sum_{nm \in B} \left( l_{nm,t} r_{nm} + P_{nm,t} \right)  \notag\\
&\quad + P_{n,t}^{DA,Load}, \quad \forall n,t \tag{2}\\
&\quad Q_{n,g,t} + Q_{n,pv,t}^{DA} + Q_{n,w,t}^{DA} = \sum_{nm \in B} \left( l_{nm,t} x_{nm} + Q_{nm,t} \right) \notag\\
&\quad+ Q_{n,t}^{DA,Load}, \quad \forall n,t \tag{3}\\
&\quad v_{n,t} = v_{n,t} - 2 \left( r_{nm} P_{nm,t} + x_{nm} Q_{nm,t} \right)  \notag\\
&\quad+ \left( r_{nm}^2 + x_{nm}^2 \right) l_{nm,t}, \quad \forall n,m \in N, nm, t \tag{4}\\
&\quad P_{nm,t}^2 + Q_{nm,t}^2 = v_{n,t} l_{nm,t}, \quad \forall n,m \in N, nm, t \tag{5}\\
&\quad P_g^{Rd} \leq P_{g,t} - P_{g,t-1} \leq P_g^{Ru}, \quad \forall g,t \tag{6}\\
&\quad 0 \leq P_{n,t}^{LS} \leq P_{n,t}^{Load}, \quad \forall n,t \tag{7}\\
&\quad e_{g,t} \underline{P_g} \leq P_{g,t} \leq e_{g,t} \overline{P_g}, \quad \forall g,t \tag{8}\\
&\quad e_{g,t} = e_{g,t-1} + s_{g,t}^{su} - s_{g,t}^{sd}, \quad \forall g,t \tag{9}\\
&\quad -\overline{P_{nm}} \leq P_{nm,t} \leq \overline{P_{nm}}, \quad \forall nm, t \tag{10}\\
&\quad V \leq v_{n,t} \leq \overline{V}, \quad \forall n,t \tag{11}
\end{align}
where \( C_g^{su} \), \( C_g \), and \( C_{ls} \) are the costs associated with the start-up of thermal units, generation, and load shedding, respectively; \( s_g^{su} \) and \( s_g^{sd} \) denote the start-up and shut-down states for thermal unit \( g \); \( P_g \) represents the power generation of thermal unit \( g \); \( P_{n,t}^{LS} \) indicates the load shedding at bus \( n \) during time \( t \); \( P_{n,g,t} \) and \( Q_{n,g,t} \) refer to the generation input at bus \( n \) in time \( t \); \( P_{n,pv,t}^{DA} \) and \( Q_{n,pv,t}^{DA} \) denote the PV power input at bus \( n \) during the day-ahead period; \( P_{n,w,t}^{DA} \) and \( Q_{n,w,t}^{DA} \) represent the wind power input at bus \( n \) in time \( t \) during the day-ahead period; \( P_{n,t}^{DA,Load} \) and \( Q_{n,t}^{DA,Load} \) are the power demands at bus \( n \) in time \( t \) during the day-ahead period; \( v_{n,t} \) is the voltage at bus \( n \) in time \( t \); \( r_{nm} \) and \( x_{nm} \) represent the resistance and reactance of the transmission line connecting nodes \( n \) and \( m \); \( P_{nm,t} \) and \( Q_{nm,t} \) are the active and reactive power flows on branch \( nm \); \( l_{nm} \) denotes the current on the transmission line \( nm \); \( P_g^{Rd} \) and \( P_g^{Ru} \) denote the ramp-down and ramp-up limits of generator \( g \); \( \underline{P_g} \) and \( \overline{P_g} \) specify the lower and upper generation limits of generator \( g \); \( e_{g,t} \) represents the on/off status of the thermal unit; \( \underline{P_{nm}} \) and \( \overline{P_{nm}} \) represent the lower and upper power flow limits on branch \( nm \); \( \underline{V} \) and \( \overline{V} \) indicate the lower and upper voltage limits for the buses.

The foresight in DP lies in its integrated approach to decision coupling and state transitions across the planning horizon. First, the objective function accounts for costs over the full cycle, requiring dispatch decisions to consider not just immediate benefits but also future load changes, renewable energy fluctuations, and network constraints for global optimality. Second, the state transition constraints, such as generator status updates and ramping limits, link current decisions to prior states, making each choice impactful for future operations.  
\vspace{-0.4cm}

\subsection{VPP adjustment for real-time fluctuation}
Once the day-ahead UC dispatch is completed, the scheduled power generation $P_{g,t}$ and the status $e_{g,t}$ from thermal generators is set. However, real-time variability in solar power, wind power, and load demand, as expressed in equations (12)-(14), can lead to power imbalances, which can be managed through the coordination of multiple VPPs described as follows:
\vspace{-0.6cm}
\begin{align}
&\min \sum_{t=1}^T \sum_{n=1}^N V_{n,t}^D + \sum_{t=1}^T \sum_{nm \in B} B_{n,t}^D + \sum_{t=1}^T \sum_{n=1}^N C_{ls} P_{n,t}^{LS}  \notag\\
&\quad+ \sum_{t=1}^T \sum_{VPP} C_{VPP} P_{VPP,t}^{VPP} \tag{12}\\
%&\min \sum_{t=1}^T \sum_{n=1}^N \left[ \left[ V - v_{n,t} \right]^+ + \left[ v_{n,t} - \overline{V} \right]^+ \right] \tag{16}\\
\text{s.t. } &\quad P_{n,g,t} + P_{n,pv,t}^{RA} + P_{n,w,t}^{RA} = \sum_{nm \in B} \left( l_{nm,t} r_{nm} + P_{nm,t} \right) \notag\\
&\quad+ P_{n,t}^{RA,Load} - P_{n,t}^{VPP}, \quad \forall n,t \tag{13}\\
&\quad Q_{n,g,t} + Q_{n,pv,t}^{RA} + Q_{n,w,t}^{RA} = \sum_{nm \in B} \left( l_{nm,t} x_{nm} + Q_{nm,t} \right) \notag\\
&\quad+ Q_{n,t}^{RA,Load} - Q_{n,t}^{VPP}, \quad \forall n,t \tag{14}\\
&\quad (4) - (7) \tag{15}
\end{align}
where \( V_{n,t}^D \) and \( B_{n,t}^D \) are the violations of voltage and branch flow; \( C_{VPP} \) denotes the cost of VPP output; \( P_{n,t}^{VPP} \) and \( Q_{n,t}^{VPP} \) are the VPP active and reactive power outputs at bus \( n \); \( [\cdot]^+ \) denotes \( \max \{0, \cdot \} \); \( P_t^{RA,PV} \) and \( Q_t^{RT,PV} \) are the PV active and reactive power output during the real-time periods; \( P_t^{RA,W} \) and \( Q_t^{RA,W} \) represent the wind active and reactive power output during the real-time periods; \( P_t^{RA,Load} \) and \( Q_t^{RA,Load} \) are the active and reactive power demand during the real-time periods.

In this process, the costs associated with voltage violations, branch flow overloads, load shedding, and VPP output are not only aggregated and considered comprehensively for each time period but also assessed with foresight into potential future fluctuations. This approach, in which every decision accounts for its impact on future states, embodies the foresight principle inherent in dynamic programming.

\section{Quantum Markov Decision Process Model}
To leverage the advantages of both quantum computing and reinforcement learning, we aim to solve the two-stage UC problem in Section II using QRL by transforming it into a q-MDP. In this section, we first introduce how the DP problem is converted into an MDP and outline the necessary assumptions for this transformation. We then describe how the key components of the MDP, particularly state transitions over time and the reward function, are mapped into q-MDP, enabling the implementation of QRL.

\subsection{MDP Formulation}
In this subsection, we firstly re-formulate the two-stage UC to two MDP models, namely day-ahead MDP and real-time MDP, with key elements including the state set $S$, action set $A$, and reward function $R$. For the day-ahead MDP, the state set contains detailed information about the power system, as defined in equation (20). The action set $a^d$, defined in equation (21), represents the operational states of the thermal generators. The reward function $r^d$, described in equation (22), evaluates the effectiveness of the DA's actions. This function accounts for factors such as operational costs, start-up costs, and penalties for load shedding to assess overall performance.
\vspace{-0.2cm}
\begin{align}
s_t^d &= \left( P_{n,t}^{DA,Load}, Q_{n,t}^{DA,Load}, P_{n,pv,t}^{DA}, Q_{n,pv,t}^{DA}, P_{n,w,t}^{DA}, Q_{n,w,t}^{DA} \right), \notag\\
&\quad \quad \forall n,t \tag{16}\\
a_t^d &= \left( e_{g,t} \right), \quad \forall t \tag{17}\\
r_t^d &= -\left( C_g^{su} s_{g,t}^{su} + C_g P_{g,t} + C_{ls} P_{n,t}^{LS} \right), \quad \forall n,g,t \tag{18}
\end{align}

As for real-time MDP, the real-time load demand and renewable energy generation are included in the state set, as defined in equation (27), to adjust the VPP output. The action set $a_{i,t}^r$ represents the power output controlled by VPP $i$ at the time t. The reward $r_t^r$ is the penalty costs associated with constraint violations, load shedding, and VPP adjustments.
\vspace{-0.2cm}
\begin{align}
s_t^r = &(P_{n,t}^{RA,Load}, Q_{n,t}^{RA,Load}, P_{n,t}^{RA,pv}, Q_{n,t}^{RA,pv}, P_{n,t}^{RA,w}, Q_{n,t}^{RA,w}  \notag\\
&\quad , e_{g,t} ), \quad \forall n, g, t \tag{19}\\
a_{i,t}^r &= ( P_{n,t}^{VPP} ), \quad \forall t \tag{20}\\
r_t^r &= -( V_{n,t}^D + B_{n,t}^D + C_{ls} P_{n,t}^{LS} + C_{VPP} P_{VPP,t}^{VPP} ), \quad \forall n, VPP, t \tag{21}
\end{align}

The transformation from the DP (in Section II) to MDP formulated above is based on several weak assumptions. First, we assume that at any given time, the system's state (e.g., day-ahead load forecasts, solar and wind generation predictions, node voltages, branch flows, and generator startup/shutdown states) fully captures all the information necessary for the system’s operation. This ensures that the future state evolution depends only on the current state and the action taken at that time, and not on past historical data. Second, it is assumed that the total cost (or total reward) of the entire system can be decomposed into the sum of the instantaneous costs for each time period, allowing the MDP model's instantaneous reward function to reflect the global optimization objective. 

\subsection{Encoding Classical Data to Quantum Bits}
In this subsection, in order to transform the elements of MDP into q-MDP and leverage quantum algorithms, the very first step is to encode the classical data to quantum bits. Here, we start from state and action data, which can be directly encoded based on their amplitude value, because the state and action data do not involve time-step transitions. For instance, a quantum state representing a vector of load measurements can be expressed in $N$-dimensional Hibert space as follows:
\begin{align}
|K\rangle = c_0 |00 \cdots 00\rangle + c_1 |00 \cdots 01\rangle + \cdots + c_{N-1} |11 \cdots 11\rangle \tag{22}
\end{align}
where each amplitude \(c_i\) is derived from the corresponding classical load value \(L_i\) at the \(i\)-th bus, where each classical data value \(L_i\) is mapped to the amplitude \(c_i\) of a corresponding computational basis state with $N$ qubits according to the rule
\begin{align}
c_i = \frac{L_i}{\sqrt{\sum_{j=0}^{N-1} |L_j|^2}} \tag{23}
\end{align}
which ensures that the resulting quantum state satisfies the normalization condition
\begin{align}
\sum_{i=0}^{N-1} |c_i|^2 = 1 \tag{24}
\end{align}
This encoding technique can be applied to encode all relevant classical data, including the system states and actions in our UC problem. 

%特别地，考虑到目前量子计算硬件的限制，对于通用的量子计算机而言，可用的量子比特数量并不多（可能只有100个左右或更少）。因此，并不能将每一个经典数据都一一映射到单独的量子比特上。

Importantly, given the current limitations of quantum computing hardware, the number of available quantum bits on general-purpose quantum computers is relatively small (typically around 100 or fewer). As a result, it is not feasible to map each classical data point to an individual quantum bit.  To address the challenge, we propose the use of a neural network linear layer with \( N \) inputs and \( M \) outputs is employed to reduce the dimensionality of the input data. The transformation performed by the linear layer is given by:
   \[
   y_j = \sum_{i=1}^{N} w_{ij} L_i + b_j, \quad \text{for } j = 1, 2, \dots, m \tag{25}
   \]
where \( y_j \) is the \( j \)-th output, \( w_{ij} \) is the weight for the \( i \)-th input and \( j \)-th output, and \( b_j \) is the bias for the \( j \)-th output. To satisfy the normalization condition required by quantum states, ensure that the \( m \) outputs are normalized: $\sum_{j=1}^{m} |y_j|^2 = 1$. This is achieved using a softmax or similar normalization layer applied to the outputs of the linear layer. 

\subsection{Quantum Markov decision process}
In this subsection, we focus on the trickier part: how to convert the state transition and reward function, with time-step transition, to the quantum state. This challenge is addressed by leveraging the density operators and quantum channels.

%Inspired by quantum computing, classical MDPs in the two-stage robust UC framework can be extended into the framework of quantum MDPs (q-MDPs). This extension leverages the inherent parallelism and probabilistic nature of quantum computing, making it possible to address complex decision-making problems that are computationally expensive or infeasible with classical MDPs. However, the construction of q-MDPs introduces several challenges that require innovative quantum approaches. To tackle these challenges, the q-MDP framework provides a robust and structured method for leveraging quantum computing in decision-making by sequentially generalizing classical MDP components through the use of density operators, quantum channels, quantum policies, and quantum measurements.

Firstly, the classical MDP probability distributions $\rho(\cdot|s,a)$, which represent the probability of transitioning to the next power system state given a state-action pair $(s,a)$, are replaced by density operators. The use of density operators is motivated by the need to represent quantum states, which, unlike classical states, can exist in superpositions and encode richer probabilistic information. Let $K(\mathcal{H})$ denote the set of density operators on a separable complex Hilbert space $\mathcal{H}$. Additionally, $\mathcal{H}_s$ and $\mathcal{H}_a$ represent separable complex Hilbert spaces representing the quantum state and action spaces, which may be infinite-dimensional. Hence, the q-MDP is formally defined by the following tuple:
\vspace{-0.2cm}
\begin{align}
(K(\mathcal{H}_s),K(\mathcal{H}_s \bigotimes \mathcal{H}_a),\mathcal{N},r) \tag{26} 
\end{align}
where the symbol $\bigotimes$ indicating a tensor product. 

Given the states and actions represented by density operators on a Hilbert space rather than by simple vectors, we further define the reward function \( R \) as follows:
\begin{align}
R&\left(K(\mathcal{H}_s \bigotimes \mathcal{H}_a)\right) = \notag\\
&\mathrm{Tr}(r \cdot K(\mathcal{H}_s \bigotimes \mathcal{H}_a)) \mathrel{\mathop:}= \langle r, K(\mathcal{H}_s \bigotimes \mathcal{H}_a) \rangle \tag{27}
\end{align}
where \(\text{Tr}(\cdot)\) is the trace operator, and $\langle r, K(\mathcal{H}_s \bigotimes \mathcal{H}_a) \rangle$ denotes the Hilbert-Schmidt inner product. 

Since states are represented by density operators on a Hilbert space, their transitions cannot be described by simple probabilistic transitions $\rho(\cdot|s,a)$ as in a classic MDP. Instead, quantum states transition under completely positive, trace-preserving maps, commonly referred to as quantum channels. These quantum channels generalize classical stochastic kernels and ensure that state transitions adhere to the fundamental principles of quantum mechanics. Accordingly, the state transition dynamics of the power system, from the state-action pair \((s_t, a_t)\) to the future state \(s_{t+1}\), as defined in Eq. (37), are also governed by quantum channels:
\vspace{-0.2cm}
\begin{align}
\mathcal{N}: K(\mathcal{H}_{s_{t}} \bigotimes \mathcal{H}_{a_t}) \rightarrow K(\mathcal{H}_{s_{t+1}})  \tag{28}
\end{align}
This quantum channel $\mathcal{N}$ encodes the Markovian property, meaning the evolution of the quantum state depends solely on the present state and action, not on past states. 

Moreover, the distribution of state transitions depends on the policy adopted by the agent. However, in quantum computing, since the state cannot be directly observed without measurement, a classical stochastic policy cannot deterministically or probabilistically select actions based on the current state. Instead, decision-making must rely on measurement outcomes or indirectly inferred information, requiring a fundamental shift in how policies are formulated in quantum environments. The introduction of the quantum policies provides a natural extension of classical policies by accounting for the probabilistic nature of quantum state transition. An admissible quantum policy is formally defined as:
\begin{small}
\begin{align}
\mathcal{X}_t :
\left(
\bigotimes_{i=0}^{t} K(\mathcal{H}_{s_i})
\right)
\otimes
\left(
\bigotimes_{i=0}^{t-1} K(\mathcal{H}_{s_i} \otimes \mathcal{H}_{a_i})
\right)
\to
K(\mathcal{H}_{s_t} \otimes \mathcal{H}_{a_t}) \tag{29}
\end{align}
\end{small}
\hspace{-0.9em}where $\mathcal{X}_t$ maps the history $K(\mathcal{H}_{s_0}) \otimes K(\mathcal{H}_{s_0} \otimes \mathcal{H}_{a_0}) \otimes \cdots \otimes K(\mathcal{H}_{s_{t-1}}) \otimes K(\mathcal{H}_{s_{t-1}} \otimes \mathcal{H}_{a_{t-1}}) \otimes K(\mathcal{H}_{s_t}) $ to an action $K(\mathcal{H}_{s_t} \otimes \mathcal{H}_{a_t})$ at time $t$. The set of all admissible policies is denoted by $\Gamma$. A policy is termed Markov if it is a sequence of quantum channels $\mathcal{X} = {\mathcal{X}_t}$, where each channel maps from the current state $K(\mathcal{H}_{s_t})$ to the corresponding action space $K(\mathcal{H}_{s_t} \otimes \mathcal{H}_{a_t})$. The set of all Markov policies is denoted by $\Gamma_m$. Furthermore, a Markov policy $\mathcal{X}$ is called stationary if $\mathcal{X}_t = \mathcal{X}_{t'}$ for all $t$ and $t'$. This step is crucial because quantum policies guide decision-making under uncertainty, allowing for the optimization of actions in complex, dynamic environments.

Under this framework, the evolution process of UC problem states and actions from time-step $t$ to $t+1$ is described by:
\vspace{-0.2cm}
\begin{align}
K(\mathcal{H}_{s_0}) = K(\mathcal{H}_{s_0}), \quad K(\mathcal{H}_{s_{t+1}}) = \mathcal{N}(K(\mathcal{H}_{s_t} \bigotimes \mathcal{H}_{a_t})) \tag{30}
\end{align}
\vspace{-0.4cm}
\begin{align}
K(\mathcal{H}_{s_t} \otimes & \mathcal{H}_{a_t}) = \gamma_t \left( K(\mathcal{H}_{s_0}) \right) \otimes K(\mathcal{H}_{s_0} \otimes \mathcal{H}_{a_0}) \otimes \cdots \tag{31}\\
&\otimes K(\mathcal{H}_{s_{t-1}}) \otimes K(\mathcal{H}_{s_{t-1}} \otimes \mathcal{H}_{a_{t-1}}) \otimes K(\mathcal{H}_{s_t})  \notag
\end{align}
This relationship highlights how the state-action mapping progresses through time, where each state transition depends on the output of a quantum channel.

\section{Quantum-based Deep Reinforcement Learning}
%In this section, we explore the application of quantum computing techniques to enhance RL algorithms to solve the formulated q-MDP. Specifically, we focus on two quantum-based algorithms: the quantum deep Q-network (Q-DQN) algorithm for addressing the day-ahead UC problem, and the quantum-based soft actor-critic (Q-SAC) algorithm for real-time optimization.
In this section, we explore the application of quantum computing techniques to enhance RL algorithms for solving the formulated q-MDP. The Q-DQN algorithm is adopted to address the day-ahead UC problem, as deep Q-networks are designed for discrete control. This characteristic makes them well-suited for the dispatch control space of the day-ahead UC problem, enabling effective solutions for discrete decision-making tasks \cite{DQN}. The quantum-enhanced version of DQN leverages quantum parallelism to accelerate learning and improve decision-making efficiency. Meanwhile, the Q-SAC algorithm is employed for real-time optimization, where its strong performance in continuous control tasks ensures stability and sample efficiency \cite{SAC}. The integration of quantum computing further enhances Q-SAC by improving exploration, reducing convergence time, and enabling more efficient policy learning in real-time optimization.

%To overcome the inherent limitations of classical DRL in addressing large-scale and dynamic UC problems, we propose transforming classical DRL into QRL, as shown in Fig.1 (right-hand side.) QRL harnesses the unique properties of quantum computing, such as superposition and entanglement, to explore the solution space more comprehensively, effectively mitigating DRL's sensitivity to local minima and saddle points. Furthermore, QRL reduces dependence on pre-existing data, accelerates convergence, and achieves robust optimization under uncertainty. These advantages enable QRL to provide a scalable, efficient, and adaptive solution that surpasses the computational and modeling constraints of DRL, particularly in modern power systems with high renewable energy penetration.

\subsection{Quantum computing based DQN algorithm}
Traditionally, the day-ahead UC, which is a discrete DP problem, can be addressed using the DQN algorithm as a classical RL method. DQN aims to learn Q-values—numerical estimates that represent the expected economic reward of the power system's operation for a given state-action pair. This process is formulated as Q-function and is shown below:
\vspace{-0.2cm}
\begin{align}
Q_{\pi}(r_t^d, a_t^d) = \mathbb{E}_{\pi} \left[ R_t \mid S_t = s_t^d, A_t = a_t^d \right] \tag{32}
\end{align}
where $R_t=\sum_{k=0}^{\infty} \gamma^k r_k $ is total reward over a sequence of steps; $\gamma$ is a discount factor; $A_t$ and $S_t$ are action and state spaces, respectively. 
The Q-function is estimated using an artificial neural network, denoted as $Q_\varphi$, where $\varphi$ represents the network's parameters. The neural network receives a power system state $s_t^d \in S_t$ as input and calculates $Q(s_t^d,a_t^d) \in S_t$ for every thermal generator control action $a_t^d \in A_t$ as its output. In order to maintain adequate exploration within a Q-network framework, an $\epsilon$-greedy policy is applied. In other words, with a probability of $\epsilon$, a random thermal generator action is taken, while with a probability of $1-\epsilon$, the agent selects the thermal generator action associated with the highest Q-value for the given power system state. At the beginning of training, $\epsilon$ is commonly chosen with a value close to 1 and gradually reduced toward 0 as learning progresses. This optimal action-value function leads to an optimal policy for selecting action $a_t^d$, where the decision-making process can be understood as a (stochastic) policy:
\vspace{-0.2cm}
\begin{align}
\pi^*(a_t^d \mid s_t^d) = \arg \max_a Q^*(s_t^d, a_t^d) \tag{33}
\end{align}
where $\sum_{a_t^d \in A_t } \pi(a_t^d|s_t^d )=1$ holds for all $s_t^d\in S_t$. 

The quantumization of the above DQN is achieved by utilizing a PQC as its Q-function approximator. The PQC consists of a sequence of unitary quantum gates, denoted as \( U(\theta, s) \), which operate on the computational base of \( n \) qubits, initialized as \( \vert 0 \rangle^{\otimes n} \). The parameters \( \theta \) control the tunable gates, while the classical input state \( s \), consisting of \( N \) components, \( s = \{ s_n \}_{n=1}^N \), is mapped into quantum states. The mapping process, \( S: \{ s_n \}_{n=1}^N \rightarrow \{ \vert K_n \rangle \}_{n=1}^N \), encodes each classical input \( s_n \) into a quantum state \( \vert K_n \rangle \). The encoding uses a rotation operation: \( S_{s_n} = \bigotimes_{n=1}^N RX(\tanh(s_n)) \), where \( RX(\cdot) \) represents a rotation around the x-axis of the Bloch sphere by an angle proportional to \( \tanh(s_n) \). 

%The PQC architecture is specifically designed to enable the representation of high-dimensional functions while capturing complex interdependencies among the input states. Each PQC layer consists of three primary components, data embedding layer, parameterized rotation gates, and entanglement layer \cite{Benedetti_2019}. 
%The data embedding layer maps the classical input \( s \) into quantum states using the aforementioned rotation operation \( RX(\tanh(s_n)) \). This embedding leverages the hyperbolic tangent function to ensure smooth and bounded input-to-state mappings.
%Following the data embedding, a set of tunable single-qubit gates \( S_{\theta_{g}} \) is applied. These gates are parameterized by learnable weights \( \theta_g \), allowing the circuit to adaptively transform the quantum state based on training data.
%To model the correlations between different input components, controlled-Z (CZ) gates are introduced. These gates create entanglement among qubits, enabling the circuit to capture joint features of the input states effectively.

The embedding process in each layer maps the classical state into the quantum space and applies unitary quantum operations to transform the state. This is mathematically defined as:
\begin{small}
\begin{align}
& U(\theta, s) =  \bigotimes_{g=1}^G \bigotimes_{n=1}^N \left( S_{\theta_{g}} \right) \notag \\
 & \left( \prod_{n=1}^N CZ(K_2 \mid K_1) \otimes \cdots \otimes CZ(K_N \mid K_{N-1}) \right) \left( S_(s_n) \right)H \tag{34}
\end{align}
\end{small}
\hspace{-0.5em}where \( H \) acting as the Hadamard gate to place a qubit in a superposition of \( \lvert 0 \rangle \) and \( \lvert 1 \rangle \) states. The operation \( \text{CZ}(\cdot) \) is a controlled-Z gate and establishes entanglement between qubits. Additionally, the quantum gate \( S_{\theta_g} \), parameterized by the initial weight \( \theta_g \) in layer \( g \), performs a rotational transformation on the qubit. Specifically, this rotation occurs around the y-axis and is represented by the \( RY(\cdot) \) operation as follows:
\begin{align}
S_{\theta_{g}} = \bigotimes_{g=1}^G RY(\tanh(\theta_{g})) \tag{35}
\end{align}

%The architecture of the PQC is chosen to balance expressivity and trainability. The inclusion of entanglement ensures that the circuit can capture correlations between inputs, while the layered structure allows incremental learning of the desired Q-function. Furthermore, the bounded nature of the rotation gates mitigates potential issues with exploding gradients during training, improving the stability and convergence of the algorithm.

To evaluate the performance of the PQC, quantum measurements are performed at the output layer to extract classical information. This output is then used to approximate the Q-values of the quantum agent based on the given input state \( s \), which is defined as follows:
\begin{align}
Q_\theta(s, a) = \langle 0^{\otimes n} | U_{\theta}^{\dagger}(s) Z  U_{\theta}(s) | 0^{\otimes n} \rangle \tag{36}
\end{align}
where the quantum observable \( Z \) is based on the Pauli-Z operator. The overall structure of an example quantum circuit architecture is shown in Fig. 3.

\begin{figure}
	\centering
	\includegraphics[width=0.5\textwidth]{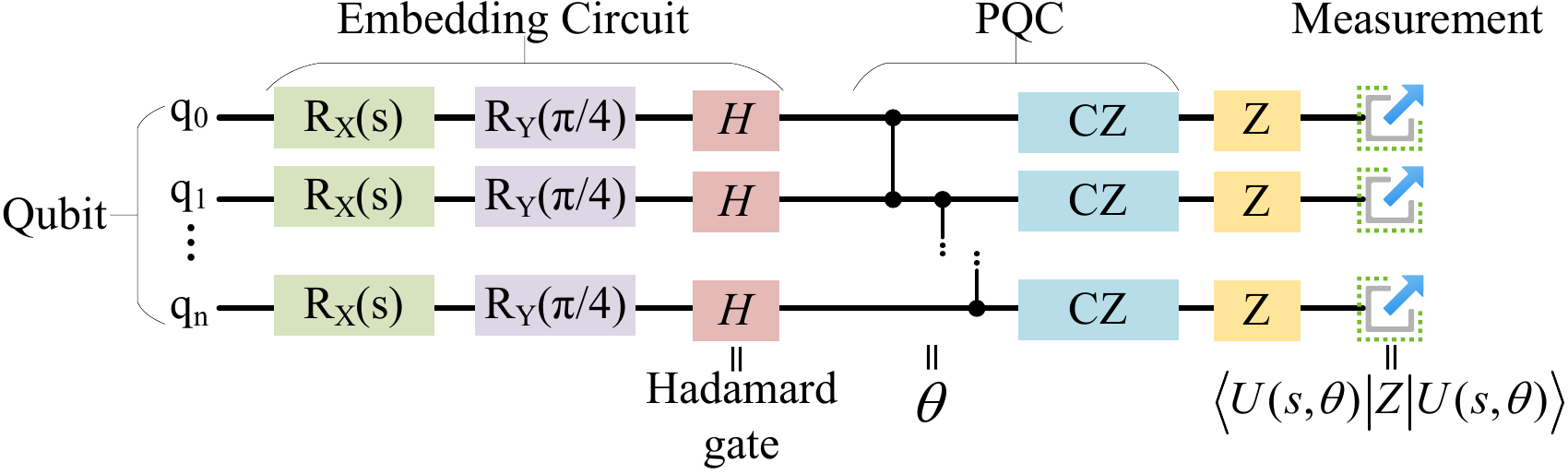} \vspace{-8mm}
	\caption{Quantum Circuit Architecture for Reinforcement Learning.}
	\label{Fig}
 	\vspace{-0.8cm}
\end{figure}
The architecture begins with the embedding circuit, where classical input data is encoded into quantum states through a sequence of rotations (\( R_X(s) \), \( R_Y(\pi/4) \)), followed by entanglement operations, controlled-Z gates. The PQC processes these quantum states through tunable gates controlled by parameters \(\theta\), capturing complex correlations among the input states. The process concludes with a measurement step, where quantum state outputs are measured to compute the expectation value of the Pauli-Z operator, reflecting the final Q-value estimate.

Moreover, after measuring the Q-value through the PQC, the Q-values need to be updated based on future states from the environment. This update rule provides direct feedback from the environment in form of the observed reward, while simultaneously incorporating the agent’s own expectation of future rewards at the present time step via the maximum achievable expected return in state $s_{t+1}$. Therefore, the Q-values are adjusted using the following update rule:
\vspace{-0.2cm}
\begin{align}
Q_\theta(r_t^d, a_t^d) \leftarrow & Q_\theta(s_t^d, a_t^d) +\alpha [ r_{t+1}^d +  \notag\\
&\gamma \cdot \max_a Q_\theta(s_{t+1}^d, a_{t+1}^d) - Q_\theta(s_t^d, a_t^d) ] \tag{37}
\end{align}
Then a target PQC, parameterized by $\theta'$, is applied to enhance the efficiency of training. This PQC shares the same architecture as $Q_\theta$ and is periodically updated, where $\theta'$ is set to match $\theta$. During each update step, a batch of $\mathcal{B}$ past experiences $(s_t^d,a_t^d,r_t^d,s_{t+1}^d)$ is sampled from the replay buffer. This batch is used to update the parameters $\theta$ using stochastic gradient descent by minimizing the difference between the target Q-value and learning Q-value:
\vspace{-0.2cm}
\begin{align}
\mathcal{L}(\theta) = \frac{1}{B} \sum \left( y_t - Q_{\theta}(s_t^d, a_t^d) \right)^2 \tag{38}
\end{align}
where $y_t=r_{t+1}^d+\gamma Q_{\theta'} (s_{t+1}^d, \text{argmax}_{{a_{t}^{d}}'} (s_{t+1},{a_{t}^{d}}')$; ${a_{t}^d}'$ is target action generated by the target PQC.
\hspace{-1.5em}

%As a result, an update is now applied to both Q-values following the update rule outlined in equation (42):
%\vspace{-0.2cm}
%\begin{align}
%Q(s_t, a_t) \leftarrow r_{t+1} + \arg \max_a Q(s_{t+1}, a) \tag{42}
%\end{align}

By leveraging quantum computing mechanisms, the Q-DQN algorithm exhibits better capability in handling high-dimensional state spaces and discrete control actions, while significantly enhancing the efficiency and effectiveness of learning optimal solutions for the day-ahead UC dispatch problem.

\subsection{Quantum computing based SAC algorithm}
This real-time optimization problem can be addressed using the SAC algorithm, an off-policy actor-critic method within the framework of maximum entropy reinforcement learning [36]. This algorithm simultaneously maximizes the expected reward and the policy's entropy, promoting better exploration based on an optimal policy as defined below:
\begin{align}
\pi^* = \arg \max_{\pi} \mathbb{E}_{\tau \sim \pi} \sum_{t=0}^{T-1} \gamma^t \left[ r + \alpha \mathcal{H} \left( \pi(\cdot \mid s_t^r) \right) \right] \tag{39}
\end{align}
where $\tau$ represents the trajectory within one episode; $\pi(\cdot \mid s_t^r)$ represents a categorical distribution that describes the probability of selecting a VPP adjustment action given the real-time state $s_t^r$. The parameter $\alpha$ controls the entropy temperature, which adjusts the level of randomness in the optimal policy. The term $H(\pi(\cdot \mid s_t^r)) = -\log(\pi(a_{i,t}^r \mid s_t^r))$ refers to the entropy component. The SAC algorithm operates within an actor-critic framework, which updates optimal policies based on maximum entropy by alternating between two networks: critic network (policy evaluation) and actor network (policy improvement). 
To improve the policy exploration, the PQC is applied in the SAC algorithm to replace the classical actor network. This adaptation involves designing a quantum-based actor network where the feed-forward learning operation is performed by the quantum circuit  $U_{\text{actor}}(\theta, s_n^{\text{actor}})$, as described in equation (46). In this case, the actor network receives $s_n^{\text{actor}} \leftarrow s(t)$ as input, representing the agent's state at time $t$. The quantum circuit $U_{\text{actor}}(\theta, s_n^{\text{actor}})$ maps the classical input $s_n^{\text{actor}}$ into the quantum space and applies a sequence of quantum operations: 
\begin{small}
\begin{align}
& U_{\text{actor}}(\theta, s_n^{\text{actor}}) = \ \bigotimes_{g=1}^G \bigotimes_{n=1}^{N_{\text{input}}^{\text{actor}}} (S_{\theta_g}) \tag{40}\\
& \left( \prod_{n=1}^{N_{\text{input}}^{\text{actor}}} CZ(K_2 \mid K_1) \otimes \cdots \otimes CZ\left( K_{N_{\text{input}}^{\text{actor}}} \mid K_{N_{\text{input}}^{\text{actor}}-1} \right) \right) (S_{s_n^{\text{actor}}}) H \notag 
\end{align}
\end{small}
%Following this, a quantum measurement is conducted at the end of each quantum gate layer, and it is represented as below:
%\vspace{-0.2cm}
%\begin{align}
%J_s(\theta_g) = Z(\theta_n) \tag{44}
%\end{align}
where $S_{s_n^{\text{actor}}}$ encodes the classical state $s_n^{\text{actor}}$ into a quantum state. This operation transforms the input into a quantum representation that facilitates policy exploration by leveraging quantum entanglement and superposition properties. The output of this quantum circuit forms the basis for generating the agent’s policy.
The policy generated by the quantum actor network is expressed as:
\begin{align}
\pi(t) = \langle 0^{\otimes n} | {U_{\theta}^{\text{actor}}}^{\dagger}(s) Z U_{\theta}^{\text{actor}}(s) | 0^{\otimes n} \rangle \tag{41}
\end{align}
where ${U_{\theta}^{\text{actor}}}$ is quantum circuit representing the actor network. Furthermore, the critic network takes in both real-time states and VPP adjustment actions as input and generates the action value $Q_{\phi}(s_t^r,a_{i,t}^r) = \mathbb{E}_{(s_t^r,a_{i,t}^r) \sim \pi} \left[ r(s_t^r,a_{i,t}^r) + \gamma V_{\pi}(s_{t+1}^r) \right]$, where $\phi$ represents the parameters of the critic network. Here, $V_{\pi}(s_{t+1}^r) = \mathbb{E}_{a_{i,t}^r \sim \pi} \left[ Q_{\phi}(s_t^r,a_{i,t}^r) - \alpha \log(\pi(a_{i,t}^r \mid s_t^r)) \right]$ defines the soft state-value function. The SAC algorithm improves the critic using temporal-difference (TD) learning by reducing an error function. To avoid overestimation of real-time robust power system dispatch, two distinct critic networks with different parameters, $\phi_1$ and $\phi_2$, are applied. Therefore, during the policy evaluation phase, the critic networks are updated using the following error function:

\vspace{-0.6cm}
\begin{align}
& J_{Q}(\phi_i) = \mathbb{E}_{(s_t^r,a_{i,t}^r,r_t^r,s_{t+1}^r) \sim \mathcal{B}} \Bigg[ \frac{1}{2} \Big( Q_{\phi_i}(s_t^r, a_{i,t}^r) - \Big( r + \tag{42}\\
& \gamma \, Q_{\hat{\phi}_i}(s_{t+1}^r, a_{i,t+1}^r) \notag  + \alpha \mathcal{H} \left( \pi^{*}(\cdot \mid s_{t+1}^r) \right) \Big) \Big)^2 \Bigg], \quad \forall i \in \{1, 2\} \notag
\end{align}
\vspace{-0.1cm}

This function calculates the TD error for the critic network and updates its parameters to minimize overestimation errors. The use of two critic networks $\phi_1$ and $\phi_2$ ensures stability and robustness.
%\vspace{-0.2cm}
%\begin{align}
%\pi(t) = \frac{1}{K_{\text{shot}}} \sum_{k=1}^{K_{\text{shot}}} Z\left(|\pi^{\theta}(t)\rangle\right) \tag{45}
%\end{align}

Notably, the objective of the actor update in Q-SAC is to maximize the expected Q-value while minimizing the logarithm of the policy, weighted by an entropy coefficient $\alpha$. This trade-off ensures that the policy not only exploits high-value actions but also retains sufficient stochasticity to effectively explore the action space. The policy gradient, used to update the quantum circuit parameters, is defined as:

\vspace{-0.6cm}

\begin{small}
\begin{align}
& J_{\text{policy}} = \mathbb{E}_{s_t^r \sim \mathcal{B}} \notag\\
& \Bigg[ \mathbb{E}_{a_{i,t}^r \sim \pi} \Big[ \alpha \log (\pi^{\theta}(a_{i,t}^r \mid s_t^r)) 
- \min_{i \in \{1,2\}} Q_{\phi_i}(s_t^r, a_{i,t}^r)   \Big] \Bigg] \tag{43}
\end{align}
\end{small}

\noindent This equation defines the policy gradient objective for the actor network. The aim is to maximize the expected Q-value (encouraging actions that yield high returns) while regularizing the policy via the entropy term $\alpha \log(\pi)$ to promote exploration. This forms the theoretical basis for optimizing the parameters $\theta$ of the PQC.

Unlike classical deep neural networks, PQCs do not allow direct gradient computation through backpropagation. Instead, the parameter-shift rule is commonly employed to compute gradients for PQCs. This rule leverages the periodic structure of quantum gates to estimate gradients by evaluating the circuit at shifted parameter values. For the quantum actor network, the gradient of $J_{\text{policy}}$ with respect to a specific parameter $\theta_k$ is expressed as:
\vspace{-0.2cm}
\begin{align}
 \frac{\partial J_{\text{policy}}}{\partial \theta_k} = \frac{1}{2} \left[ J_{\text{policy}}\left(\theta_k + \frac{\pi}{2}\right) - J_{\text{policy}}\left(\theta_k - \frac{\pi}{2}\right) \right] \tag{44}
\end{align}

\noindent The parameter-shift rule computes the gradient by symmetrically perturbing the parameter $\theta_k$ around its current value ($\theta_k + \frac{\pi}{2}$ and $\theta_k - \frac{\pi}{2}$). Here, $J_{\text{policy}}(\theta_k + \frac{\pi}{2})$ and $J_{\text{policy}}(\theta_k - \frac{\pi}{2})$ represent the policy objective evaluated at these shifted parameter values. This approach requires evaluating the PQC twice for each parameter, making it computationally practical for quantum hardware.

The parameter-shift rule provides an exact gradient for commonly used quantum gates, such as rotation gates ($RX, RY, RZ$). It offers a practical means to estimate the gradient of the PQC by evaluating the objective function $J_{\text{policy}}$ at shifted parameter values.

Once the gradients $\frac{\partial J_{\text{policy}}}{\partial \theta_k}$ are computed for all parameters $\theta_k$, the PQC parameters are updated using a gradient-descent-like rule:
\vspace{-0.3cm}
\begin{align}
\theta_k \leftarrow \theta_k - \eta \cdot \frac{\partial J_{\text{policy}}}{\partial \theta_k} \tag{45}
\end{align}

\noindent where $\eta$ is the learning rate.

%The loss for each layer can subsequently be computed as below:
%\vspace{-0.2cm}
%\begin{align}
%\mathcal{L}(\theta^{\pi}) = \frac{1}{N_{\text{data}}} \sum_{i=1}^{N_{\text{data}}} - \left( h_{\text{actor}}(s_i(t)), h_{\text{actor}}(s_i(t)) \theta^{\pi}(t-1) \right) \tag{46}
%\end{align}
%where $\mathcal{L}(\theta^{\pi})$ represents the training loss of the policy network, and $h_{\text{actor}}(\cdot)$ indicates the quantum circuit function for the policy network. Using the computed loss $\mathcal{L}(\theta^{\pi})$ and referencing equation (47), the gradient step for the actor network at the output layer can be determined as below:
%\vspace{-0.2cm}
%\begin{align}
%\nabla_{\theta^{\pi}} \mathcal{L}(\theta^{\pi}(t-1)) = \frac{1}{2} \left( \mathcal{L}(\theta^{\pi}(t-1)+ \delta) - \mathcal{L}(\theta^{\pi}(t-1) - \delta) \right) \tag{47}
%\end{align}

%Subsequently, the parameters of the actor network are updated as below:
%\vspace{-0.2cm}
%\begin{align}
%\theta^{\pi}(t) = \theta^{\pi}(t-1) - \eta_{\text{actor}} \nabla_{\theta^{\pi}} \mathcal{L}(\theta^{\pi}(t-1)) \tag{48}
%\end{align}
%where parameter $\eta_{\text{actor}}$ indicates the learning rate of the policy network.
%These improvements enable Q-SAC to achieve faster and more effective learning of optimal solutions, highlighting its potential to address real-time decision-making challenges in dynamic robust UC problem.
\vspace{-0.2cm}
\section{Case study}
In this section, a modified IEEE RTS 24-bus system is utilized to evaluate the stability improvements achieved by the robust UC model and the performance of the QRL algorithm. The system comprises 11 thermal generators, two solar farms located at buses \#1 and \#22, and two onshore wind farms connected to buses \#2 and \#23. To enhance system reliability, improve power quality, and reduce real-time transmission losses, eight VPPs  are installed at buses \#3, \#5, \#6, \#8, \#9, \#14, \#17, and \#20. The structure of the IEEE RTS 24-bus system is illustrated in Fig. 4. 
%Seven consecutive days of historical load demand data ($24 \times 7$ loads) are presented in Fig. 5. To account for all possible load conditions, a set \( U = \{ u_t \mid u_t \in [(\bar{u}_t - 3\sigma_t), (\bar{u}_t + 3\sigma_t)] \} \) is defined. Load data are randomly sampled from this set to serve as states for the reinforcement learning model. 

\begin{figure}
	\centering
	\includegraphics[width=0.45\textwidth]{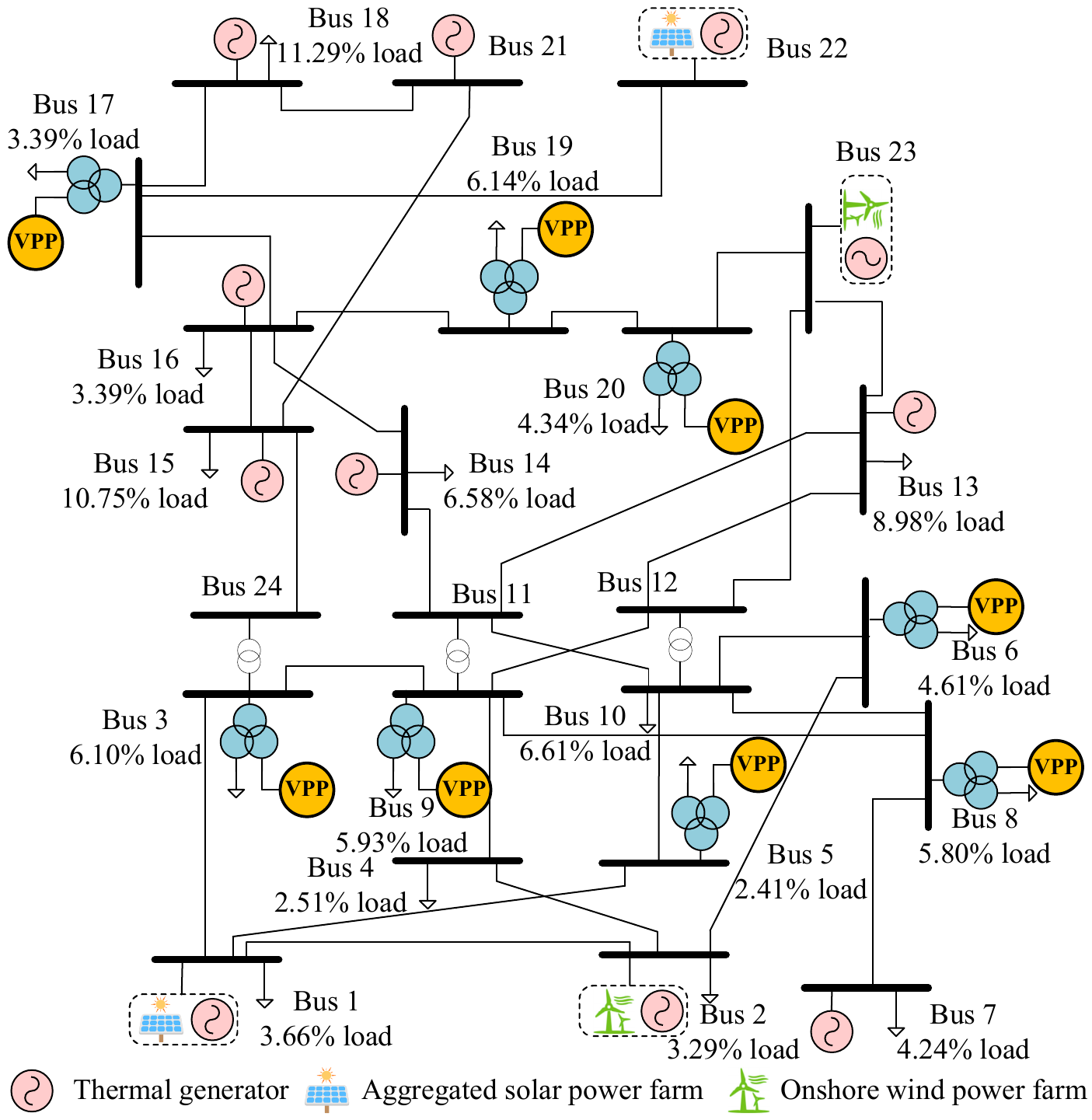} % Use textwidth for spanning both columns
	\vspace{-4mm}
	\caption{Structure of modified IEEE RTS 24-bus system.}
	\label{fig:4}
	\vspace{-0.6cm}
\end{figure}

\subsection{Comparative Analysis of QRL in Solving the Robust UC Problem}
This section of the simulation compares the performance of four optimization methods in solving the robust UC problem, focusing on the dispatch of various energy sources. The four approaches are linear programming (LP), relaxed linear programming (RLP), DRL, and QRL. The results are averaged over seven days and presented to show the energy output over 24 hours for each method, as illustrated in Fig. 6 and summarized in Table II.

\begin{figure}
	\centering
	\includegraphics[width=0.45\textwidth]{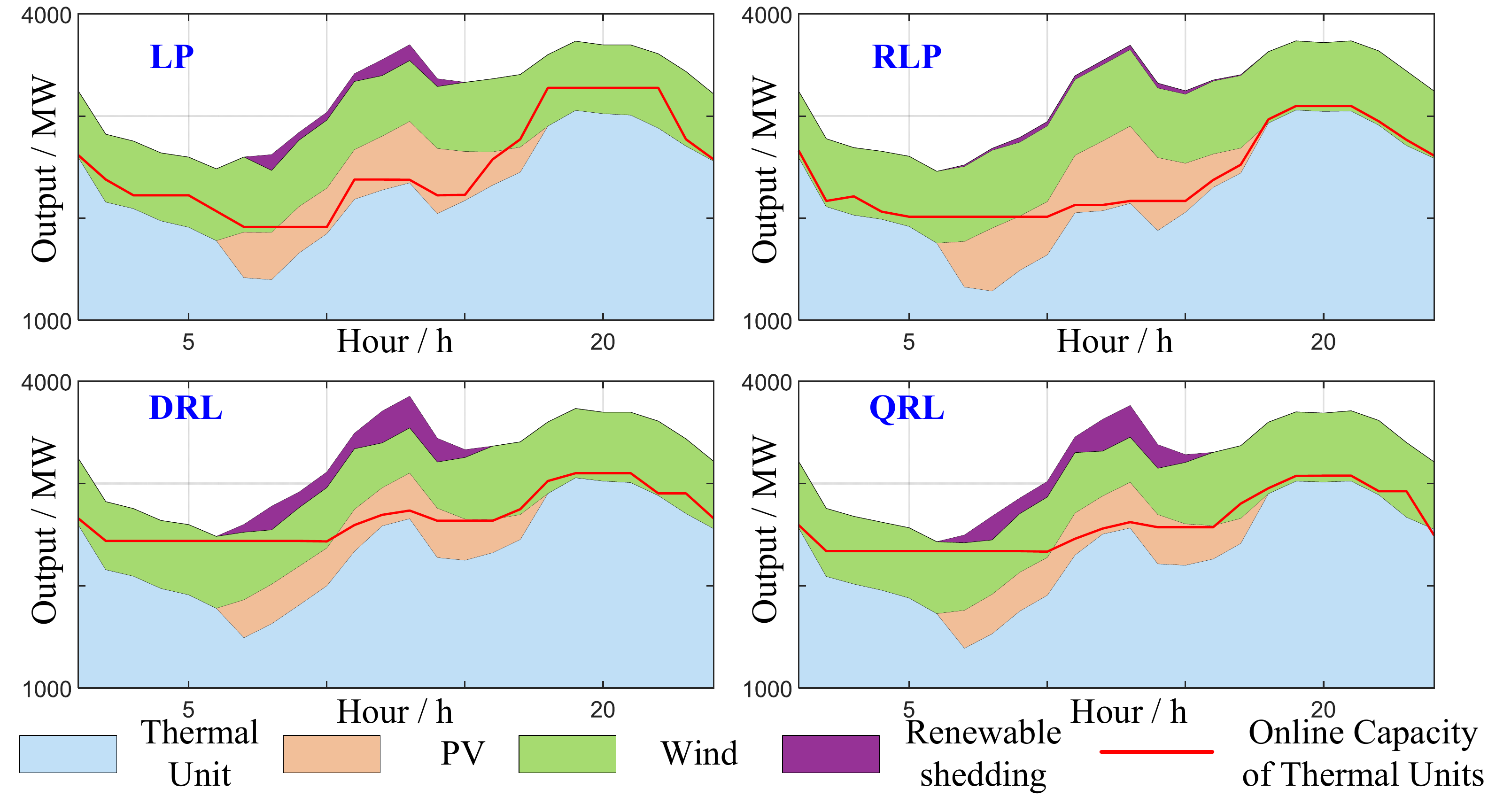} \vspace{-4mm}
	\caption{Power generation stack of four different optimization approaches.}
	\label{fig:4}
 	\vspace{-0.4cm}
\end{figure}

\begin{table}[!ht]	
	\centering
	\caption{Comparison results of four different optimization approaches}\label{tab:1}
	\vspace{-0.25cm}
	\begin{tabular}{c|cccc}
		\toprule
		& LP & RLP & DRL & QRL\\ \midrule
		Toal Cost   & 0.7948&	0.7581&	0.7501&	0.7506\\
		Operational cost   & 0.4948&	0.4881&	0.5201&	0.5106 \\
		Start-up cost  & 0.3&	0.27&	0.23&	0.24 \\
		Violation & 0&	0.4983 p.u.&	0 0546 p.u.&	0. 0118 p.u.\\
		Time & 76.70s&	1.14s&	1.628s&	2.617s \\
		\bottomrule
	\end{tabular}
\end{table}

The results of the four optimization methods—LP, RLP, DRL, and QRL—show notable differences in solving the robust UC problem. LP incurs the highest total (0.7948) and operational costs (0.4948) due to its heavy reliance on thermal units, resulting in high start-up costs (0.3) and longer computation time (76.70s). However, it has no violations, ensuring strict adherence to constraints. RLP reduces total (0.7581) and operational costs (0.4881) and start-up costs (0.27) by increasing renewable energy usage, but this results in higher violations (0.4983 p.u.).
DRL further lowers costs with a total of 0.7501 and the lowest start-up cost (0.23), while being more computationally efficient (1.628s). However, it leads to increased renewable energy shedding. QRL achieves the best balance, reducing total (0.7506) and operational costs (0.5106) with low start-up costs (0.24) and the lowest violation rate (0.0118 p.u.), optimizing both costs and constraint compliance. While its runtime is slightly longer (2.617s), QRL offers the best overall performance.

Fig. 6 shows that the DRL method outperforms the traditional LP approach in a seven-day UC dispatch scenario, achieving lower total, fuel, and UC costs. This advantage comes from DRL's ability to anticipate future rewards through episodic learning, optimizing long-term benefits. In contrast, LP optimizes dispatch daily without considering future states, leading to higher costs. DRL's dynamic decision-making allows for better adaptation to load and renewable generation fluctuations, improving resource management and reducing operational costs.

%\textcolor{blue}{QRL's ability to satisfy constraints may come from two quantum properties. First, quantum amplitude encoding allows constraints to be embedded in quantum states, reducing violations through interference effects. Second, quantum entanglement between qubits may help represent interactions between different generators more directly, making it easier to find solutions that work within multiple constraints. This could explain our observation that QRL has fewer violations (0.0118 p.u.) than classical methods. As systems get larger, this quantum approach seems more beneficial since quantum states can handle multiple constraints simultaneously through interference.}

\begin{figure}
	\centering
	\includegraphics[width=0.45\textwidth]{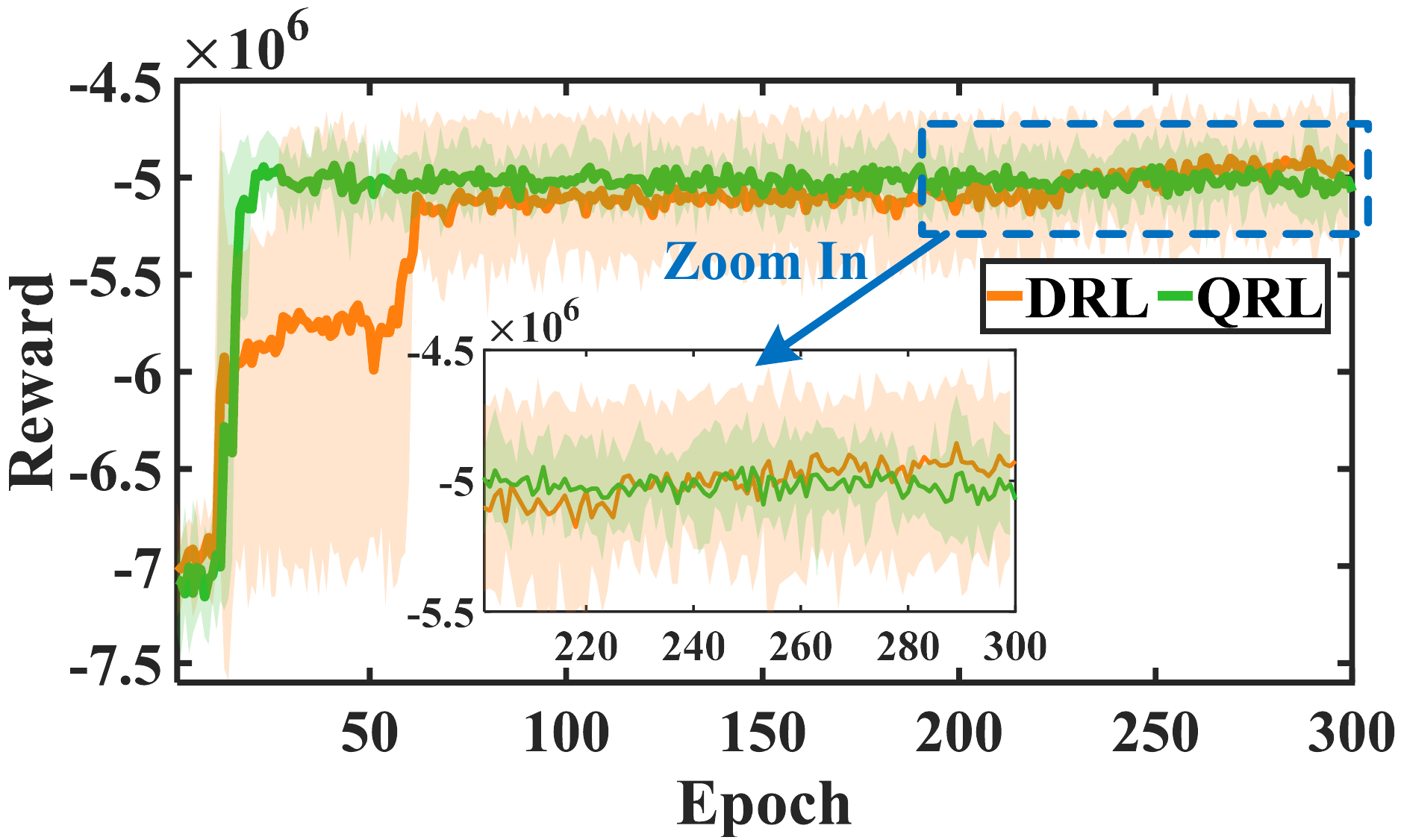} \vspace{-4mm}
	\caption{Comparison results of operational cost between QRL and DRL approaches.}
	\label{fig:4}
 	\vspace{-0.2cm}
\end{figure}

\subsection{Comparison Analysis with DRL algorithm}

This section compares the performance of QRL and DRL in solving the robust UC problem, tested in ten independent runs with different initial seeds. Fig. 8 shows the cumulative reward curves, where each solid curve represents the average value, and the shaded area indicates the range of rewards. Table III summarizes the training outcomes, highlighting differences in dispatch results, violation levels, and adjustment outputs between QRL and DRL.

\begin{figure}
	\centering
	\includegraphics[width=0.45\textwidth]{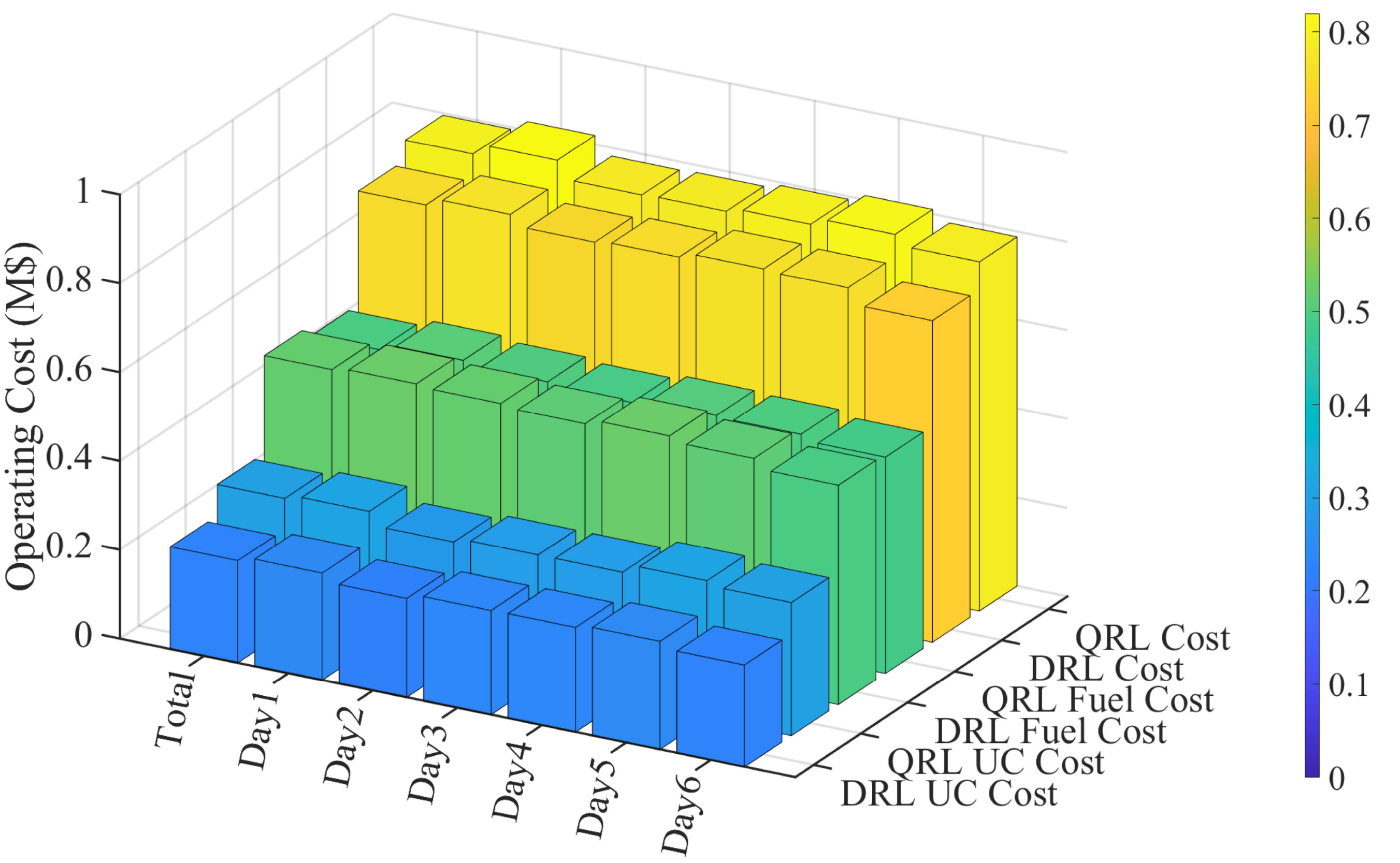} \vspace{-4mm}
	\caption{Training process of different algorithms.}
	\label{fig:4}
 	\vspace{-0.4cm}
\end{figure}

\begin{table}[!ht]	
	\centering
	\caption{Training results with different algorithms}\label{tab:1}
	\vspace{-0.25cm}
	\begin{tabular}{c|ccc}
		\toprule
		& Reward & Violation degree & Adjustment output\\ \midrule
		DRL  & -4.9530&	0.6892&	7728.97MW\\
		QRL  &-5.0192&	0.0883&	8268.88MW \\
		\bottomrule
	\end{tabular}
\end{table}

The figures show that QRL converges faster, reaching a steady reward of around -5.0 by epoch 50, while DRL takes longer and exhibits more variability across 300 epochs. Between epochs 200 and 300, QRL remains more consistent, while DRL fluctuates more. DRL achieves a slightly better reward (-4.9530) than QRL (-5.0192), but with significantly higher violation levels (0.6892 vs. 0.0883), suggesting DRL sacrifices constraint compliance for a better reward. Additionally, QRL has a higher adjustment output (8268.88 MW) compared to DRL (7728.97 MW), indicating QRL is more active in maintaining stability. In conclusion, QRL converges faster and better controls system violations, making it more suitable for applications requiring strict system constraints. 

\begin{figure}
	\centering
	\includegraphics[width=0.45\textwidth]{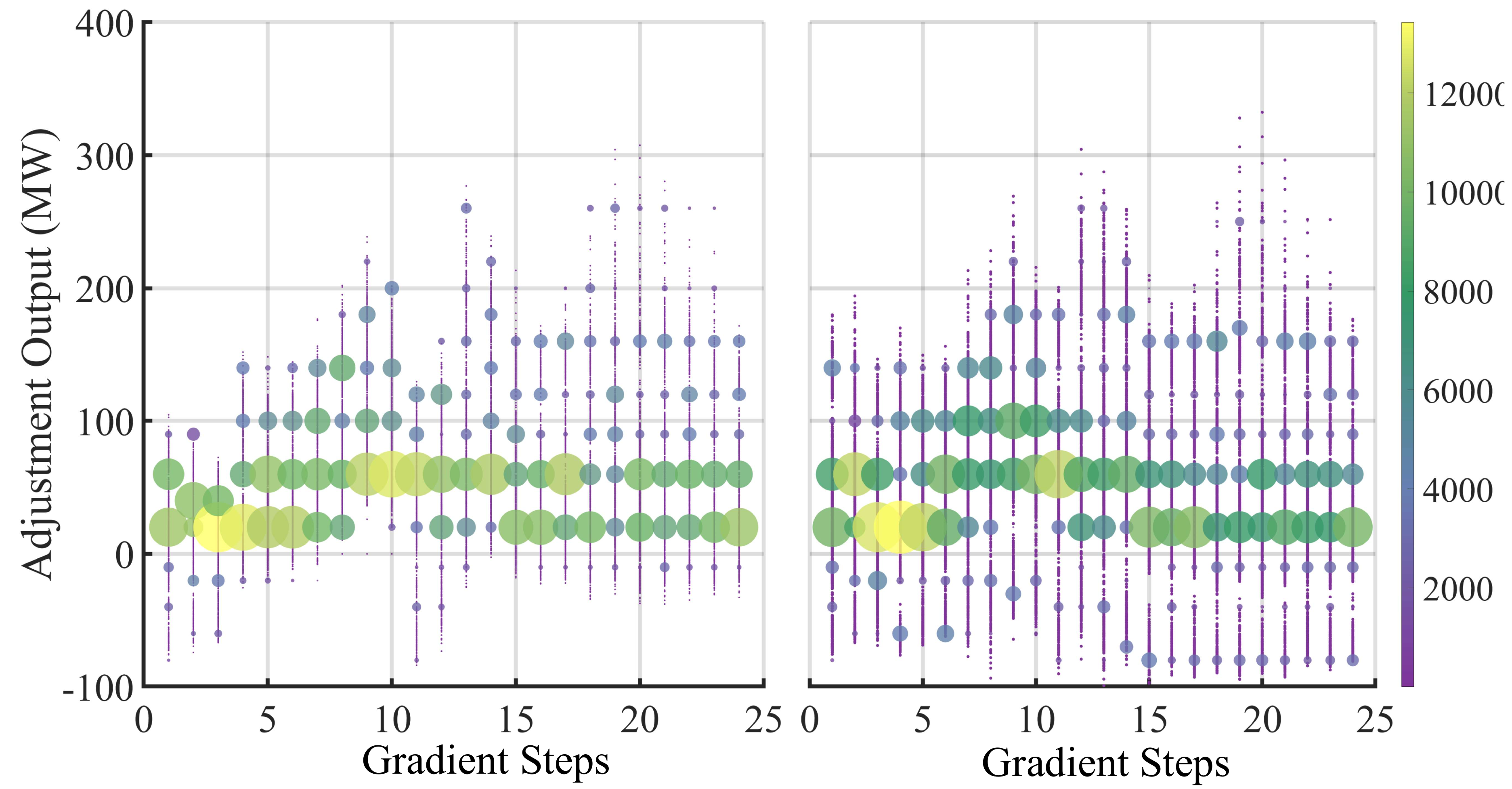} \vspace{-4mm}
	\caption{Distribution of VPP adjustment with different algorithms.}
	\label{fig:4}
 	\vspace{-0.6cm}
\end{figure}

Fig. 9 compares the VPP adjustment outputs controlled by DRL (left) and QRL (right), highlighting key differences in how each algorithm maintains power system stability. DRL shows a more concentrated and less varied adjustment distribution, indicating a focus on a narrower set of solutions, which limits its flexibility under fluctuating conditions. In contrast, QRL displays a broader and more varied distribution, benefiting from quantum computing techniques like superposition and entanglement to explore a wider solution space. This allows QRL to search for optimal solutions more effectively.
QRL’s broader distribution helps avoid local optima, a common challenge for DRL. By evaluating multiple optimization paths simultaneously, QRL is better at finding the global optimum, making it more robust for real-time VPP adjustments in complex scenarios. Therefore, QRL’s wider solution distribution underscores its superiority in solving the robust UC problem, providing more efficient solutions compared to DRL.

\begin{figure}
	\centering
	\includegraphics[width=0.5\textwidth]{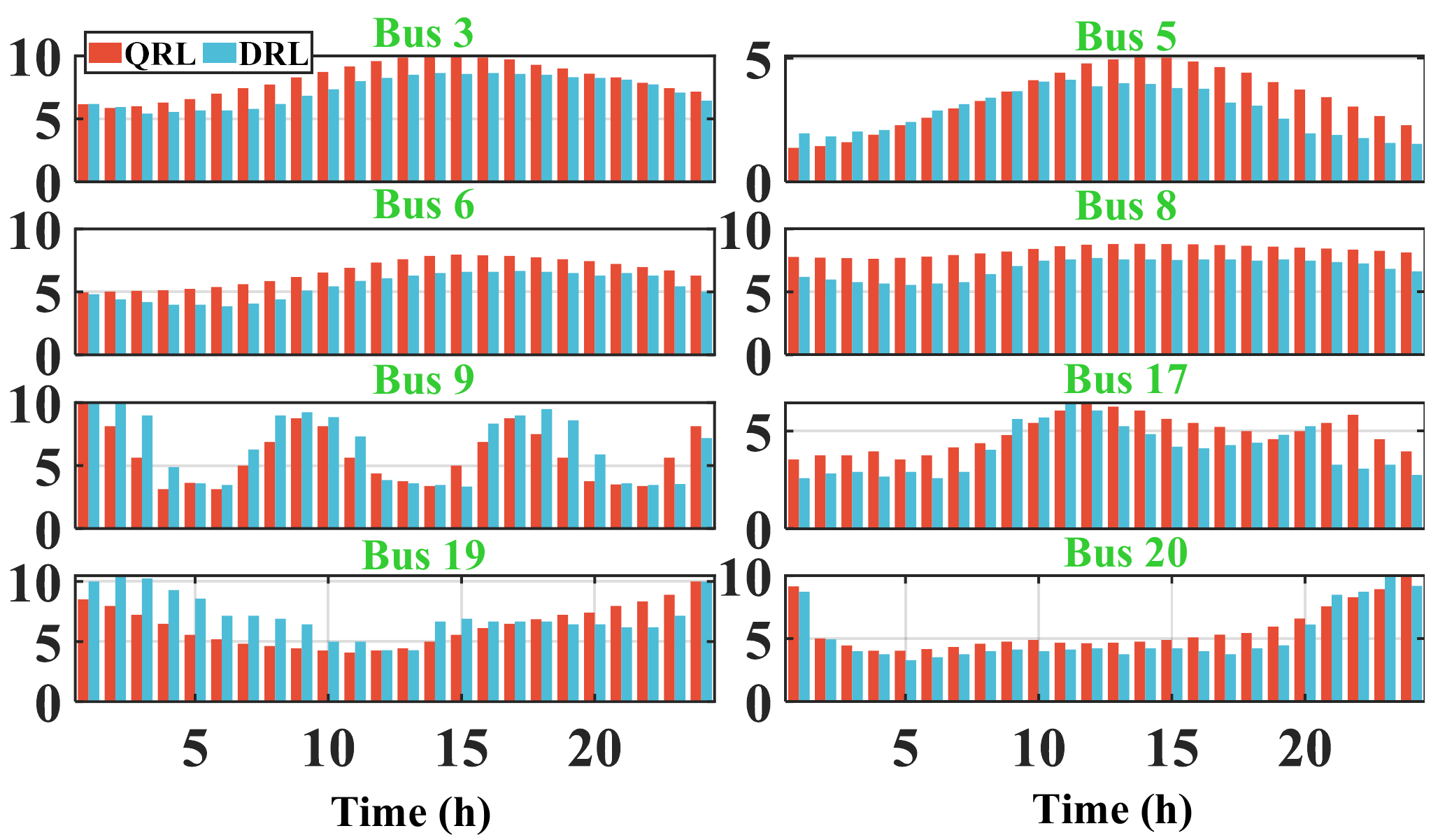} \vspace{-4mm}
	\caption{Detail value of VPP adjustment with different algorithms.}
	\label{fig:4}
 	\vspace{-0.6cm}
\end{figure}

Fig. 10 compares the VPP adjustment outputs of QRL and DRL across eight subplots, showing QRL's larger, more dynamic adjustments (red bars) versus DRL's smaller, more conservative ones (blue bars). In VPPs like VPP 1 and VPP 2, QRL produces higher adjustment peaks, especially during peak times, while DRL maintains minimal adjustments. This trend is consistent across all VPPs, with QRL responding more actively to higher demand, especially in VPP 3 through VPP 8, while DRL shows less variability. The performance difference is clear: QRL’s quantum-enhanced exploration allows for more effective optimization, dynamically adjusting to real-time fluctuations. In contrast, DRL’s classical approach limits significant adjustments, making it less responsive. As a result, QRL is more efficient and robust in complex optimization scenarios, while DRL offers stable but less flexible responses.

\section{Conclusion}
This study introduces a novel two-stage UC framework designed to enhance the robustness and responsiveness of power systems in the face of increasing uncertainty from renewable energy sources and load fluctuations. By incorporating QRL into the UC process, the proposed approach effectively addresses the limitations of conventional methods, which often suffer from computational inefficiency and infeasibility under real-time conditions. A key contribution of this work is the use of QRL to handle the complexity and high dimensionality of the UC problem without relying on approximations or constraint relaxations. This significantly improves both computational efficiency and the reliability of real-time dispatch decisions. The integration of VPPs further demonstrates the potential of localized resources to support system-wide stability under uncertain conditions. Validation on the IEEE RTS 24-bus system confirms the framework’s effectiveness, offering a viable alternative to traditional UC methods. The improvements in performance, particularly in dynamic real-time scenarios, highlight the potential of QRL for broader applications in power system optimization.

\bibliographystyle{IEEEtran}

\bibliography{IEEEabrv, ref.bib}

\end{document}